\documentclass[conference]{IEEEtran}
\IEEEoverridecommandlockouts

\usepackage{cite}
\usepackage{amsmath,amssymb,amsfonts}
\usepackage{graphicx}
\usepackage{textcomp}
\usepackage{xcolor}
\usepackage{url}

\begin{document}

\title{Hardware-Transparent I/O Governance in Disaggregated Heterogeneous Storage}

% IC2E is single-blind: replace placeholder block below with author names and affiliations.
\author{\IEEEauthorblockN{Rajarshi Chowdhury, Akshay Shah, Sue K. Lee}
\IEEEauthorblockA{Oracle America Inc \\
\{rajarshi.chowdhury, akshay.shah, sue.k.lee\}@oracle.com}
}

\maketitle

%-------------------------------------------------------------------------------
\begin{abstract}
%-------------------------------------------------------------------------------
Shared-nothing disaggregated storage clusters that serve both latency-sensitive databases and opaque block-volume workloads face two governance problems unsolved by existing schedulers: maintaining consistent performance across heterogeneous hardware generations, and enforcing global I/O limits when access patterns skew to a subset of storage nodes. We present the I/O Resource Manager (IORM), a multi-stage distributed scheduler deployed in production within Oracle Exadata Exascale. IORM combines three mechanisms: a hardware-aware cost modeler that normalizes I/O accounting using datasheet-derived fixed costs to make limits invariant across hardware generations; a quantum-based rate limiter with bounded carry-forward credits that accommodates database micro-bursts while enforcing long-term SLOs; and a distributed adaptive feedback controller that redistributes unused entitlements across the cluster to resolve topological access skew. Beyond design, we share operational lessons from production deployment. On an 8-node test cluster running up to 100 concurrent tenant volumes, IORM converges within 5\% of provisioned limits under extreme sequential skew, scales without inter-tenant interference, and recovers full throughput within 15 seconds of a storage-node failure.
\end{abstract}


%===============================================================================
% SECTION 1 — INTRODUCTION (~0.75 pages)
%===============================================================================
\section{Introduction}

Hyperscale cloud storage increasingly relies on disaggregated, shared-nothing architectures in which compute nodes (database instances, virtual machines) are decoupled from storage nodes and interconnected via a high-speed Remote Direct Memory Access (RDMA) network fabric~\cite{klimovic2016flash, aguilera2017remote, shan2018legoos}. Physical disks across all storage nodes are aggregated into a global Storage Pool that is carved into logical tenancies. This pool is persistent, shared, and heterogeneous: rolling hardware upgrades mean that multiple storage-server generations coexist within a single cluster. Two fundamentally different I/O classes compete for this shared pool. \emph{Structured database I/O} carries semantic tags (e.g., transaction-log flush vs.\ analytics scan) enabling priority-aware scheduling~\cite{cao2020rocksdb}, while \emph{unstructured block-volume I/O} from guest VMs is semantically opaque and often exhibits severe access skew.

Each tenant---whether a distributed database or a set of block volumes---is provisioned with a global IOPS limit that must be enforced across the entire cluster. This is a fundamentally harder problem than per-node rate limiting. On the compute side, a distributed database or clustered application spans multiple compute nodes, each independently issuing I/O to the storage layer. A Linux cgroup controller~\cite{axboe2004linux} on any one compute node can only throttle the I/O originating from that node; it has no visibility into the aggregate I/O the tenant is generating across all compute nodes. On the storage side, each storage node only observes the I/O arriving at its local disks---it cannot know how much of the tenant's budget is being consumed at peer storage nodes. Cgroup-based enforcement on either side is therefore insufficient: the workload is distributed across compute nodes, the data is striped across storage nodes, and no single vantage point sees the tenant's total consumption. What is needed is a \emph{distributed} scheduler that coordinates across all storage nodes to enforce a single logical limit while remaining transparent to the physical hardware underneath.

Two unsolved challenges make this difficult. The first is the \textbf{Hardware Gap}. The physical capability of NVMe media evolves faster than software provisioning policies, and large clusters routinely host several hardware generations side by side as rolling upgrades replace nodes incrementally rather than wholesale. In Exadata Exascale, for example, an X9M storage server (rated for roughly 2.3 million 8\,KB Read IOPS over PCIe~4.0) typically coexists with an X10M server (roughly 2.8 million IOPS over PCIe~5.0 with denser NAND) and a newer X11M generation in the same cluster. Standard block-layer schedulers such as CFQ, BFQ, and Kyber~\cite{valente2012bfq, axboe2004linux} measure I/O ``cost'' via wall-clock device busy time, which couples the cost they perceive to the speed of whichever device happens to be underneath. Because a modern storage server completes the same 8\,KB read substantially faster than a prior-generation one, a time-based scheduler inadvertently grants tenants on newer hardware more effective throughput for the same logical entitlement. Prior studies confirm that SSD latency and throughput vary significantly across Flash generations even under identical workload patterns~\cite{agrawal2008ssdtradeoffs, li2021latency}. The result is performance inconsistency: a tenant's observed IOPS fluctuates with whichever generation of storage server happens to serve their requests.

The second challenge is \textbf{Distributed Access Skew}. Sequential or single-threaded workloads---software installations, VDI boot storms, streaming writes---concentrate their I/O on a small subset of storage nodes while leaving the rest idle. A naive ``static slicing'' approach divides the global limit evenly across $N$ nodes ($Limit_{local} = Limit_{global}/N$). When a workload targets one node, it is throttled to $1/N$ of its entitlement while the remaining capacity sits stranded on idle nodes. Distributed schedulers such as PARDA~\cite{gulati2009parda} enforce fairness across hypervisors but lack feedback-driven credit redistribution for shared-nothing clusters. IOFlow~\cite{ioflow2013} provides path-level policy enforcement but has no notion of hardware-aware I/O cost or dynamic rebalancing.

This paper presents the I/O Resource Manager (IORM), a multi-stage distributed scheduler deployed in production within Oracle Exadata Exascale~\cite{bhatia2024exascale}, a large-scale heterogeneous cloud-storage platform. IORM runs locally on every storage node as an embedded component of the storage-server software. It makes four contributions:
\begin{enumerate}
    \item A \textbf{Hardware-Aware Cost Modeler} that normalizes I/O accounting using datasheet-derived fixed costs, making provisioned limits invariant across hardware generations (Section~\ref{sec:costmodel}).
    \item A \textbf{Quantum-Based Rate Limiter} that operates on 200\,ms control quanta with carry-forward credits and anti-windup clamping, accommodating database micro-bursts while enforcing strict long-term SLOs (Section~\ref{sec:ratelimiter}).
    \item A \textbf{Distributed Adaptive Feedback Controller} that dynamically redistributes unused I/O entitlements across the cluster using EMA smoothing, proportional gain, and slew-rate damping to resolve topological access skew without inducing oscillation (Section~\ref{sec:controller}).
    \item A set of \textbf{operational lessons} from production deployment---pathologies that motivated specific design choices, parameter trade-offs that emerged at scale, and the practical limits of static-cost modeling---intended to inform future cloud-storage governance designs (Section~\ref{sec:lessons}).
\end{enumerate}
We evaluate IORM on a controlled 8-node test cluster (mirroring the production deployment) running up to 100 concurrent tenant volumes. The system converges to within 5\% of provisioned limits under extreme sequential skew, scales without inter-tenant interference, and recovers full throughput within 15 seconds of a storage-node failure (Section~\ref{sec:eval}).

%===============================================================================
% SECTION 3 — HARDWARE-AWARE COST MODELER (~0.75 pages)
%===============================================================================
\section{Hardware-Aware Cost Modeler}\label{sec:costmodel}

To external users, storage performance is specified in IOPS (Input/Output Operations Per Second), with the industry-standard unit being a small random I/O of 4\,KB or 8\,KB. Major cloud providers define provisioned IOPS at these baseline block sizes (e.g., 4\,KB for OCI Block Volumes~\cite{ociblockperf}, 16\,KB for AWS EBS SSD volumes~\cite{awsebsio}). Larger I/O sizes consume proportionally more of the device's throughput budget, reducing the effective IOPS: a 64\,KB sequential read costs roughly 8$\times$ a single 8\,KB random read. Internally, IORM does not schedule on raw IOPS counts. Instead, it converts the user-supplied IOPS target into a \emph{utilization percentage} relative to the specific hardware generation, enabling what we call \textbf{Hardware Transparency}.

As discussed in Section~I, time-based schedulers conflate physical device speed with logical cost. To make the magnitude of this concrete, consider a tenant provisioned at 500{,}000 IOPS in a mixed-generation cluster. On an X9M storage server, rated for roughly 2.3 million 8\,KB Read IOPS, that provisioning works out to about 22\% of total capacity. On the next-generation X10M, rated for roughly 2.8 million IOPS thanks to PCIe~5.0 interfaces and higher-density NAND, the same provisioning is only about 18\%. A time-based scheduler throttles the older server earlier and harder than the newer one, so the same tenant's effective throughput fluctuates depending on which hardware happens to serve a given request.

IORM resolves this by decoupling logical accounting from physical latency. Upon initialization, each storage server queries its hardware profile and retrieves the datasheet-rated maximum IOPS and throughput for the installed media. It then computes a static, normalized cost for three I/O classes: small reads, small writes, and large I/Os.

IORM classifies every I/O request against a configurable threshold (default: 64\,KB). Requests below this threshold are \emph{small I/Os}, governed by IOPS limits; requests at or above it are \emph{large I/Os}, governed by throughput limits. The two classes are costed against different hardware maximums because they stress different parts of the device: small random I/Os are latency-bound while large sequential I/Os are bandwidth-bound.

For small I/Os, the scheduler quantizes each request in units of the baseline block size $B$ (configurable, default 8\,KB). A request smaller than $B$ is charged a minimum of one quantum; larger small requests are charged proportionally:
\begin{equation}
    Q_{small} = \max\!\Big(\Big\lfloor \frac{IOSize}{B} \Big\rfloor,\; 1\Big)
\end{equation}
Small reads and small writes are accumulated into separate counters ($N_{small\_rd}$, $N_{small\_wr}$), since Flash read and write latencies differ significantly across media types and generations, and their datasheet-rated maximums are different. In a redundant storage configuration, writes are almost always mirrored to multiple storage nodes to ensure data protection. To avoid penalizing write-heavy tenants, the cost modeler counts only the primary copy of each write against the tenant's budget; mirror copies are excluded from utilization accounting.

For large I/Os ($\geq$\,64\,KB), the scheduler computes a throughput-normalized quantum count that scales with the payload size:
\begin{equation}
    Q_{large} = \Big\lceil \frac{IOSize \times L_{ref}}{L_{limit} \times B} \Big\rceil
\end{equation}
where $L_{ref}$ is a reference large-I/O size and $L_{limit}$ is the device's large-I/O limit granularity. Large I/Os are tracked in a single counter ($N_{large}$) regardless of read/write direction, as at these sizes the throughput bottleneck dominates over the read/write latency asymmetry.

Each hardware generation defines three datasheet-derived costs: $Cost_{small\_rd}$, $Cost_{small\_wr}$, and $Cost_{large}$, computed from the device's rated maximum small-read IOPS, small-write IOPS, and sequential throughput respectively. The total utilization consumed by a tenant in a given interval is:
\begin{multline}
    Util_{measured} = N_{small\_rd} \times Cost_{small\_rd} \\
    + N_{small\_wr} \times Cost_{small\_wr} + N_{large} \times Cost_{large}
    \label{eq:util}
\end{multline}
On a higher-capability server the per-quantum costs are smaller (larger denominators in the datasheet maximums), so the same logical workload consumes less utilization---automatically admitting proportionally more operations and exactly matching the hardware's increased physical capability.

This fixed-cost design also provides accounting consistency during Flash garbage collection: GC cycles cause internal latency spikes~\cite{hu2012writeamp}, and a time-based scheduler would penalize tenants during GC by charging them more for the same I/O. IORM's datasheet-derived costs are invariant to transient device-internal latency, ensuring that tenants are not unfairly billed for device maintenance---though the physical latency of individual I/Os during GC is still incurred at the device level.

The cost model requires a lookup table of per-generation device capabilities, populated from published datasheets or from a calibration pass during system initialization and updated when new hardware generations are introduced. While this introduces an operational maintenance cost, it is a one-time effort per generation and is amortized across all tenants and workloads in the cluster.

The static cost model is a deliberate design choice that trades accuracy under edge conditions for operational simplicity and determinism. Flash device performance can degrade over time due to write amplification (WAF) as devices fill, and the effective write cost increases sharply when the SLC write cache is exhausted and writes fall through to TLC or QLC NAND~\cite{maneas2020fastssd, yang2014dontstack}. The current model does not capture these non-linearities; it prices I/Os based on peak rated capability. In practice, this means that during sustained heavy writes that exhaust the SLC cache, the scheduler may slightly under-account the true physical cost. We found this acceptable because the cost model's primary purpose is cross-generation normalization and fair relative accounting between tenants, not precise physical modeling of device internals. The carry-forward credit system (Section~\ref{sec:ratelimiter}) and the feedback controller (Section~\ref{sec:controller}) absorb residual variance that the cost model does not capture.

%===============================================================================
% SECTION 4 — QUANTUM-BASED RATE LIMITER (~0.75 pages)
%===============================================================================
\section{Quantum-Based Rate Limiter}\label{sec:ratelimiter}

Database workloads impose conflicting scheduling requirements: strict long-term SLO enforcement and tolerance for intense sub-millisecond micro-bursts (e.g., transaction-log flushes that must complete with minimal latency). At cluster scale these latency excursions are not a curiosity; even rare tail events disproportionately shape end-to-end response time, as classic studies of large fan-out services have shown~\cite{dean2013tail}. Standard token-bucket rate limiters struggle to reconcile these goals---a deep bucket permits SLO-violating sustained bursts, while a shallow bucket clips legitimate transaction commits.

IORM resolves this tension by operating on a fixed 200\,ms control quantum. This interval is short enough to contain latency excursions within human-perceptible thresholds, yet long enough to amortize the cost of floating-point utilization accounting across thousands of concurrent consumer groups. At each quantum boundary, the limiter computes $Util_{measured}$ for each tenant using the three-class cost model from Section~\ref{sec:costmodel} (Equation~\ref{eq:util}). This measured utilization is compared against the tenant's entitlement limit ($Limit_{pct}$) scaled to the elapsed wall-clock time:
\begin{equation}
    Util_{allowed} = \Delta t \times Limit_{pct}
\end{equation}

A naive limiter would throttle whenever $Util_{measured} > Util_{allowed}$, but this clips valid transaction bursts in bursty database workloads. IORM instead implements a carry-forward credit system: if a tenant consumes less than its entitlement in quantum $t_0$, the surplus $(Util_{allowed} - Util_{measured})$ is banked. If the tenant bursts in a subsequent quantum $t_1$, it draws down this balance. Throttling engages only when the carry-forward balance becomes negative.

Unbounded credit accumulation, however, creates a ``windup'' pathology: a tenant idling for an extended period could accumulate enough credit to monopolize the disk upon waking. IORM prevents this by clamping the maximum carry-forward balance to one second of burst capacity:
\begin{equation}
    Carry_{max} = 1 \text{ second} \times Limit_{pct}
\end{equation}
This creates a bounded \emph{burst window}: a tenant may burst to 100\% of disk bandwidth to commit a transaction, but only for at most one second. Once the credit is exhausted, the limiter enforces a hard clamp back to the provisioned average, preserving the SLAs of co-located tenants. The one-second bound was selected empirically as sufficient for the largest observed transaction-commit bursts while remaining operationally safe for neighboring workloads.

Several design choices in the rate limiter were refined through production deployment. The 200\,ms quantum was arrived at iteratively: earlier designs used 100\,ms, which doubled the CPU overhead of the floating-point utilization math for thousands of concurrent consumer groups without measurably improving SLO accuracy; a 500\,ms quantum reduced CPU cost but allowed latency excursions visible to interactive database workloads. The anti-windup clamp at one second replaced an earlier unbounded design that caused a recurring production issue: idle databases that accumulated hours of credit would wake up and monopolize the disk for minutes, triggering cascading latency spikes for co-located tenants. The one-second bound eliminated this pathology while preserving enough headroom for legitimate transaction-commit bursts observed in production OLTP workloads.

%===============================================================================
% SECTION 5 — DISTRIBUTED ADAPTIVE FEEDBACK CONTROLLER (~0.75 pages)
%===============================================================================
\section{Distributed Adaptive Feedback Controller}\label{sec:controller}

\begin{figure*}[t]
\centering
\includegraphics[width=0.95\textwidth]{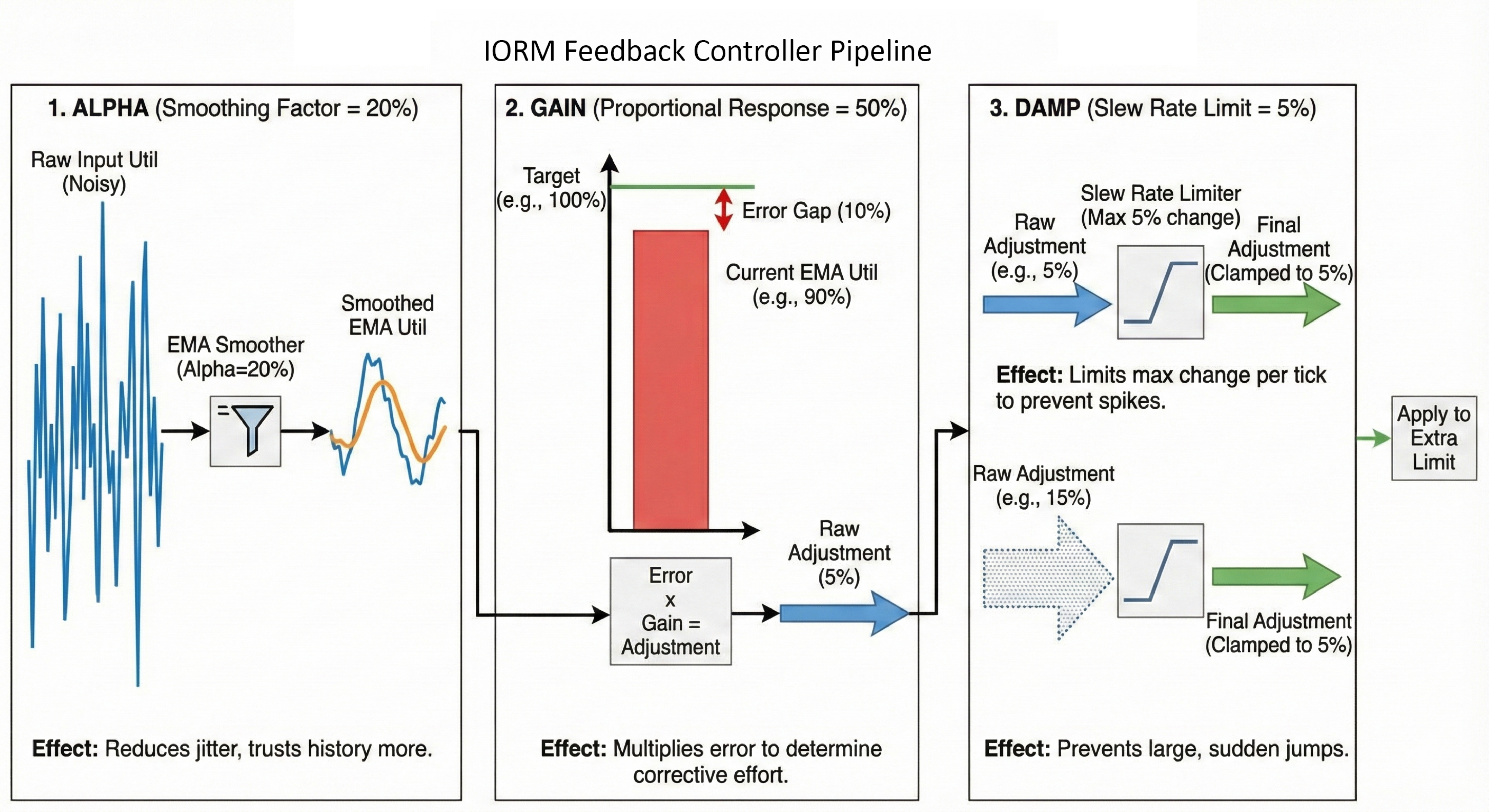}
\caption{The IORM distributed feedback controller pipeline, running independently on each storage node. Stage~1 (Alpha) smooths noisy utilization via EMA to reduce jitter. Stage~2 (Gain) computes a proportional corrective adjustment from the error gap between target and observed utilization. Stage~3 (Damp) clamps the adjustment to $\pm$5\% per tick, preventing large sudden jumps. The bounded output is applied to the node's extra limit, allowing hot nodes to borrow unused capacity from idle peers.}
\label{fig:controller}
\end{figure*}

The cost modeler and rate limiter operate locally on each storage node, but a tenant's global IOPS limit must be enforced across the entire cluster. A baseline approach is \emph{static slicing}: divide the global limit evenly ($Limit_{local} = Limit_{global} / N_{nodes}$). This fails under skewed access patterns. When a sequential workload targets one of eight storage nodes, for example, the tenant is throttled to $1/8$ of their entitlement on that node while the remaining $7/8$ sits stranded on idle peers.

IORM resolves this with a Distributed Adaptive Feedback Controller that runs locally on every storage node but coordinates cluster-wide state. Each node computes a local \emph{extra limit} adjustment that allows hot nodes to borrow unused capacity from idle nodes, effectively decoupling the logical limit from the physical topology. The controller processes utilization data through a three-stage pipeline designed to prevent cluster-wide oscillation, shown in Figure~\ref{fig:controller}:

The first stage applies an Exponential Moving Average (EMA) with smoothing factor $\alpha = 0.2$ to filter noisy utilization measurements caused by I/O bursts and device-internal variance:
\begin{equation}
    U_{ema}(t) = \alpha \cdot U_{measured}(t) + (1-\alpha) \cdot U_{ema}(t-1)
\end{equation}
This weighting prevents the controller from reacting to transient GC spikes or micro-bursts while still tracking sustained skew shifts. In the second stage, the controller computes the error between the target utilization (100\% of the global limit) and the current smoothed utilization, then applies a proportional gain:
\begin{equation}
    \varepsilon(t) = TargetUtil - U_{ema}(t)
\end{equation}
\begin{equation}
    Adj(t) = \varepsilon(t) \times K_p
\end{equation}
A gain of $K_p = 0.5$ means the system closes only half the error gap per tick. This conservative response prevents overshoot; larger gains were observed empirically to cause successive adjustments across nodes to amplify each other, producing ring-convergence oscillations characteristic of unstable distributed control loops~\cite{padala2009control, hellerstein2004feedback}. In the third stage, regardless of the computed adjustment magnitude, the controller clamps the per-tick change to $\pm$5\% of the node's capacity:
\begin{equation}
    Adj_{final} = \text{Clamp}(Adj, -0.05, +0.05)
\end{equation}
This bound ensures gradual ramp-up and prevents synchronous ``gate opening'' across the cluster---a phenomenon where multiple nodes simultaneously raise their limits, causing fabric-level congestion and thundering-herd contention.

To prevent thrashing during intermittent activity, the controller implements a hysteretic state machine. In the \emph{Active} state, the pipeline above runs continuously. When measured utilization drops to zero, the controller enters a \emph{Hold} state, preserving the current limit for a configurable grace period (default: 5\,s) to avoid cold-start penalties when a bursty workload temporarily pauses. If inactivity persists beyond the grace period, the controller transitions to \emph{Decay} mode, gradually releasing the extra limit to free capacity for other tenants.

\begin{figure*}[t]
\centering
\includegraphics[width=0.95\textwidth]{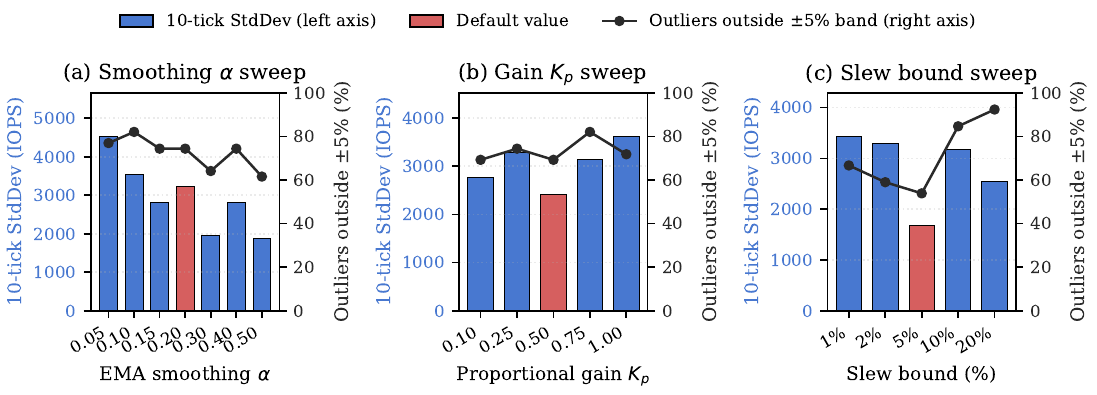}
\caption{Controller parameter sensitivity. For each of the three knobs ($\alpha$, $K_p$, slew bound), we sweep across a representative range while holding the other two at their production defaults (highlighted in red), driving a 200-tick deterministic simulation of the distributed controller (3-node configuration) against the same skewed write workload pattern used in Section~\ref{sec:eval}. Bars show the standard deviation of throughput at a 10-tick (2\,s) averaging window; the line shows the fraction of windows whose mean falls outside the $\pm$5\% provisioned-limit band. Lower is better on both axes. The slew-bound sweep (c) shows the strongest effect: outlier fraction climbs from 54\% at the 5\% default to 92\% at 20\%.}
\label{fig:sensitivity}
\end{figure*}

Under this control loop, a skewed workload targeting a single storage node converges as follows: the controller detects idle capacity on $N{-}1$ peer nodes, ramps up the hot node's extra limit at 5\% per tick via the EMA/gain pipeline, and reaches steady state within seconds. At convergence, the hot node's local ceiling is raised to permit the full global entitlement---bounded only by the physical media---while idle nodes retain minimal reserved capacity. The 5\% damping ensures smooth monotonic ramp-up rather than oscillatory convergence.

\subsection{Parameter Selection}
The three controller parameters ($\alpha = 0.2$, $K_p = 0.5$, $Slew_{max}=\pm$5\%) are exposed as runtime configurables but were tuned to defaults that have remained stable across multiple platform generations. They were not derived analytically. Instead, they emerged from observation of production behavior under workload mixes that span OLTP transaction storms, analytics scans, and unstructured VDI/streaming traffic, and were subsequently validated by the controller-level sweep we present below.

Two production observations are worth surfacing because they materially shaped the defaults. First, raising the smoothing factor $\alpha$ above $\sim$0.35 caused the controller to chase Flash GC-induced microbursts, producing visible jitter in the redistributed limit---a workload-dependent effect we discuss further in Section~\ref{sec:lessons} and which the controller-level sweep does not exercise. Second, the slew bound is the most operationally consequential of the three parameters: it directly governs the synchronous gate-opening pathology described above, and even modest increases above 5\% degrade SLA compliance significantly, as the next subsection's measurements confirm.

\subsection{Quantifying Parameter Sensitivity}
To validate these qualitative choices with measurement, we ran a controller-level sweep. For each of the three parameters in turn, we varied that parameter across a representative range while holding the other two at their production defaults and drove a 200-tick deterministic simulation of the distributed controller against a skewed write workload---the same controller code path that runs in production, exercised on a 3-node simulator under reproducible synthetic input. (The 8-node hardware test cluster used in Section~\ref{sec:eval} is not amenable to deterministic parameter sweeps because device-level variance dominates at the time scales we wish to characterize.) We report two stability metrics computed over each run: the standard deviation of throughput at a 10-tick (2\,s) averaging window, which captures sustained oscillation; and the fraction of 10-tick windows whose mean falls outside the $\pm$5\% provisioned-limit band, which captures SLA compliance. Results are shown in Figure~\ref{fig:sensitivity}.

The slew-bound sweep (Fig.~\ref{fig:sensitivity}(c)) is the cleanest result and the most operationally important. The default 5\% bound is the only setting that brings the outlier fraction below 60\%, with a 10-tick StdDev of 1{,}674 IOPS---roughly half the value at any other tested point. Above 5\%, outlier fraction climbs sharply to 85\% at 10\% and 92\% at 20\%, confirming the synchronous gate-opening pathology described above: at 20\%, every node is permitted to raise its limit by a fifth of capacity per tick, and even with independent local controllers their decisions correlate enough to overshoot the cluster-wide target on the majority of quanta. Below 5\%, ramp-up is too slow to track workload shifts and stddev rises again, though more gradually.

The proportional-gain sweep (Fig.~\ref{fig:sensitivity}(b)) shows a shallower U-shape centered on the default $K_p=0.5$, which produces the lowest StdDev (2{,}418 IOPS) and tied-best outlier rate. $K_p=1.0$ raises StdDev to 3{,}620 IOPS but does not produce the dramatic ring-convergence collapse we have observed in production at larger cluster sizes, suggesting that the pathology requires more than three nodes' worth of correlated adjustment to manifest. We retain the default as a conservative margin.

The smoothing-factor sweep (Fig.~\ref{fig:sensitivity}(a)) shows the expected sluggish-controller penalty at very low $\alpha$ (StdDev = 4{,}530 IOPS at $\alpha=0.05$, 33\% worse than the default), but does not reproduce the high-$\alpha$ degradation we observe in production. This is informative in itself: the simulated workload is bursty in topology but not in device latency---there are no Flash GC microbursts to chase, so a high-$\alpha$ controller does not over-react. The qualitative claim that $\alpha > 0.35$ causes jitter is therefore a workload-dependent observation that the controller-level test does not exercise. We surface this limitation explicitly because it reinforces a broader point: parameter tuning in distributed I/O governance must integrate both controller dynamics (visible here) and device dynamics (not), and a simulator that captures only the former will systematically understate the cost of poor smoothing choices.

In summary, the sweep validates the production parameter choices for the slew-bound and gain dimensions while honestly bounding the simulator's ability to characterize $\alpha$. The slew bound is the dominant operational lever, with outlier rate increasing nearly $2\times$ as the bound is widened from 5\% to 20\%---a stronger effect than either of the other two parameters produces in any tested setting.

%===============================================================================
% SECTION 6 — EVALUATION (~1.0 page)
%===============================================================================
\section{Evaluation}\label{sec:eval}

IORM is deployed in production across multiple large-scale disaggregated storage clusters serving diverse enterprise workloads. The controlled experiments below were conducted on a dedicated 8-node test cluster to enable reproducible workload generation and destructive fault injection that would not be possible on live production infrastructure. The test cluster consists of 8 High-Capacity Storage Servers connected via RDMA (RoCE) to Compute Nodes running multiple VMs with mixed structured and unstructured workloads. Structured traffic is generated by a production database engine using semantically rich access patterns. Unstructured traffic is generated by block volumes executing concurrent mixes of disruptive, high-skew patterns---software installations, VDI boot storms, streaming workloads, and single-threaded sequential reads and writes---to simulate realistic ``noisy neighbor'' contention.

Throughout these experiments we compare IORM against the simplest distributed alternative, which we call \emph{static slicing}: each storage node enforces $Limit_{global}/N$ on its own, with no cross-node coordination. We picked this as the baseline for two reasons. First, it is exactly what tenants get if we disable IORM's feedback controller while leaving the underlying token-bucket rate limiter in place, so it cleanly isolates what the controller adds. Second, it is the regime tenants actually experience in many production storage systems today: a per-node share of an evenly divided global budget. Richer coordinated alternatives, such as weighted-fair queuing or distributed token-bucket schemes~\cite{liu2008distributedrl}, would need their own cluster-wide coordination layer and are better viewed as variants of the design space IORM occupies than as alternatives to it. A head-to-head comparison there would mostly measure differences in coordination protocols rather than the governance abstractions this paper is about; we return to that broader design space in Section~\ref{sec:related}.

The experiments report aggregate IOPS convergence, which is the SLO our production deployment is contracted to enforce. End-to-end I/O latency is governed primarily within each control quantum by the cost-modeled token bucket described in Section~\ref{sec:ratelimiter}: tagged latency-sensitive operations such as transaction-log flushes are admitted ahead of bulk traffic and absorb only the small queueing delay introduced by the 200\,ms quantum boundary. Because the cost modeler turns every I/O into a normalized utilization charge before the rate limiter sees it, latency-class tagging survives across the redistribution path even when bursty block-volume neighbors are simultaneously trying to spend their own carry-forward credits.

\subsection{Skew Resolution}

To evaluate the controller's ability to enforce a global provisioned limit under extreme topological skew, we run a single-threaded sequential write stream (simulating a software installation). Uncapped, this workload saturates the backing devices it touches, achieving an aggregate throughput of 144,000 IOPS. We apply a logical limit of 32,000 IOPS. Because the workload is sequential, it hammers a single storage server at a time before moving to the next rather than striping evenly.

\begin{figure}[t]
\centering
\includegraphics[width=\columnwidth]{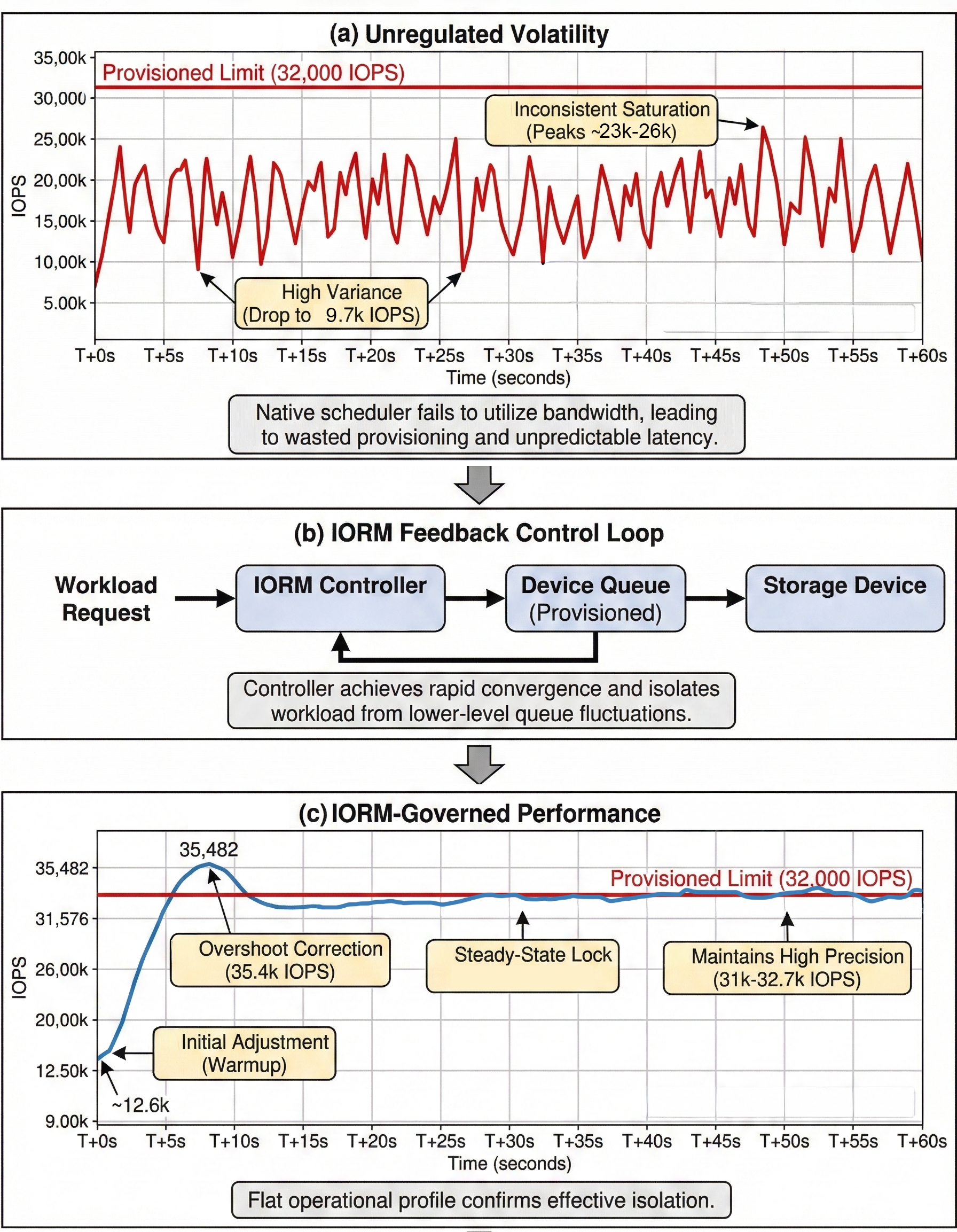}
\caption{Resolution of distributed access skew. Without IORM (top), static slicing produces sawtooth volatility between 9,700 and 26,000 IOPS. With IORM (bottom), the feedback controller detects idle capacity on peer nodes and floats credits to the active node, converging to the 32,000 IOPS target within 10 seconds.}
\label{fig:skew_results}
\end{figure}

Figure~\ref{fig:skew_results}(a) shows performance under static slicing (no feedback controller). The workload exhibits severe sawtooth volatility, oscillating between 9,700 and 26,000 IOPS and never reaching the 32,000 IOPS provisioned limit. The tenant is penalized purely for their access pattern.

Figure~\ref{fig:skew_results}(b) shows the same workload under IORM governance. The feedback controller detects unused credits on idle storage servers and dynamically floats them to the active node. After a brief initialization warmup ($T{+}0$s to $T{+}5$s) and a momentary overshoot to fill the hardware queue (35.4K IOPS), the controller locks onto the target. From $T{+}10$s onward, the workload flatlines at approximately 32,000 IOPS---within 5\% of the provisioned limit. IORM successfully decoupled the logical limit from the physical topology.

\subsection{Multi-Tenant Scaling}

To verify that the feedback controller scales without inducing inter-tenant interference, we provision 100 concurrent block volumes, each independently configured with a 32,000 IOPS limit. All 100 volumes simultaneously execute a diverse mix of workloads: database table scans, software installations, VDI boot storms, streaming workloads, and synthetic sequential reads and writes.

\begin{figure}[h]
\centering
\includegraphics[width=\columnwidth]{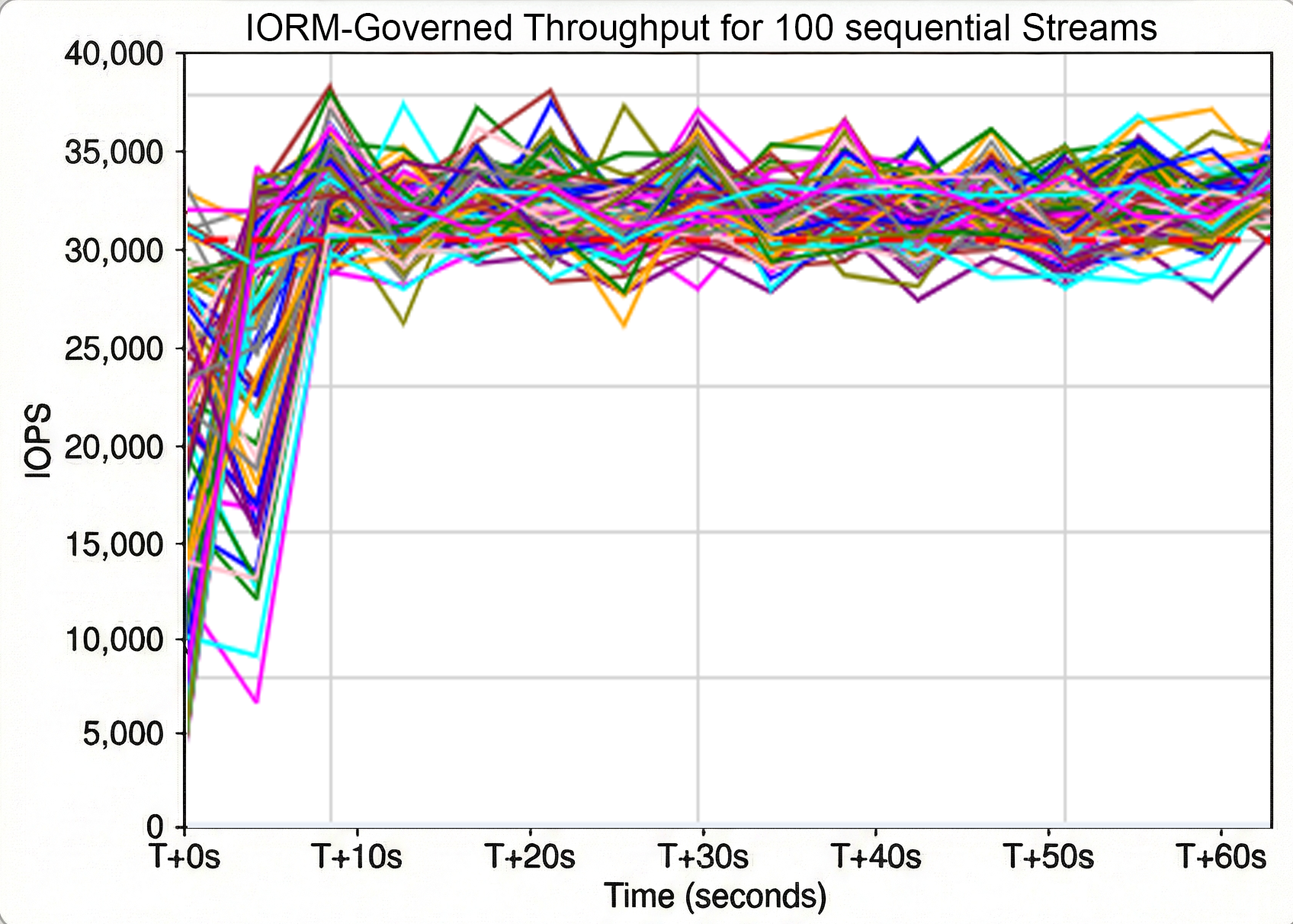}
\caption{Massively parallel governance. Throughput overlay of 100 concurrent volumes over 60 seconds. Every volume converges to its 32,000 IOPS limit within seconds with no evidence of thundering-herd instability or inter-tenant interference.}
\label{fig:concurrency}
\end{figure}

Figure~\ref{fig:concurrency} overlays the throughput traces of all 100 volumes over a 60-second interval. Despite the massive contention and 100 independent feedback loops running concurrently, the dense band of traces shows that every volume converged to its 32,000 IOPS limit within seconds. There is no evidence of thundering-herd instability or inter-tenant interference. The controller scales linearly, effectively isolating 100 concurrent noisy neighbors.

\subsection{Fault Recovery}

To evaluate robustness under degraded hardware conditions, we inject a fault during steady-state operation of the 32,000 IOPS workload. At approximately $T{+}30$s, a single storage server is forcibly offlined to simulate sudden component failure.

\begin{figure}[h]
\centering
\includegraphics[width=\columnwidth]{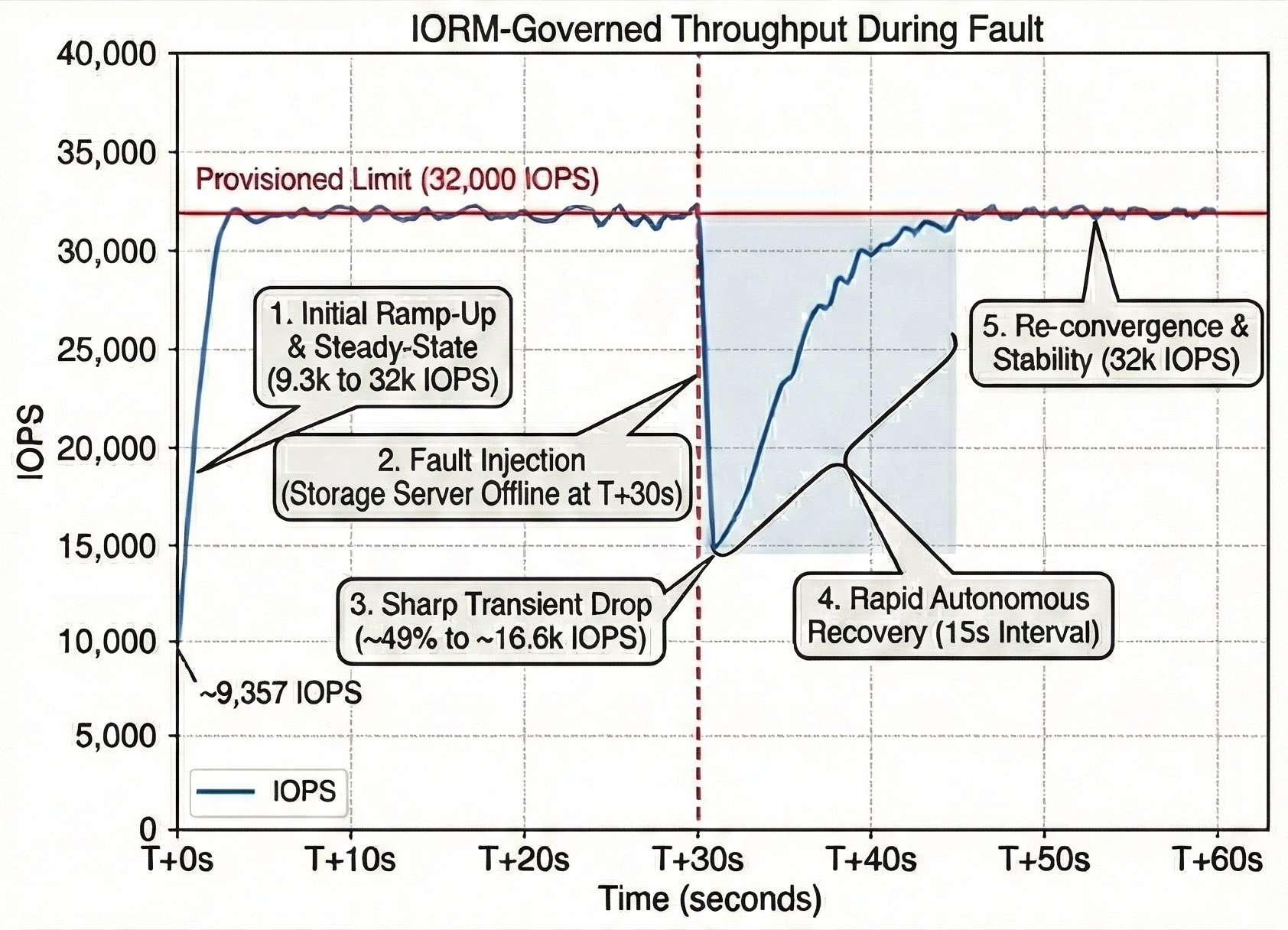}
\caption{Automatic recovery from storage server failure. At $T{+}30$s, one storage server is offlined. Throughput drops 48\% as active streams redirect to mirror copies. The feedback controller re-converges to the 32,000 IOPS setpoint within 15 seconds.}
\label{fig:fault}
\end{figure}

Prior to the fault, the system stably maintains the provisioned limit. Upon losing the storage server, throughput drops sharply by 48\%, indicating that a significant fraction of active sequential streams were targeted at the failed node and must redirect to redundant mirror copies on surviving servers. The feedback controller detects the new topology within one control cycle and begins redistributing credits to the remaining nodes. As shown in Figure~\ref{fig:fault}, the system re-converges to the 32,000 IOPS setpoint within 15 seconds. The tenant suffers no long-term performance degradation; the scheduler fully masks the hardware loss by delivering the logical entitlement across survivor nodes.

%===============================================================================
% SECTION 7 — LESSONS LEARNED (~0.75-1.0 page) -- CORE OF INDUSTRY-TRACK PAPER
%===============================================================================
\section{Lessons Learned}\label{sec:lessons}

IORM has been refined over multiple platform generations and many years of production deployment. The mechanisms in Sections~\ref{sec:costmodel}--\ref{sec:controller} look architecturally clean in retrospect, but most of what distinguishes them from a textbook design was forced by production behavior rather than analysis. In the spirit of an experience report, we now share five lessons that we believe generalize to other distributed storage governance designs.

A first lesson is that datasheet-derived costs are good enough in practice, and that the cases where they fail are best handled downstream rather than by making the cost model itself smarter. A common objection to fixed-cost modeling is that it ignores real device dynamics: write amplification grows as devices fill, the SLC write cache eventually drains into TLC or QLC NAND with several-times-higher write latency, and aging media drift below their datasheet ratings. We considered a dynamic recalibration scheme that would adjust the cost table from observed I/O latency. The recalibration loop introduced a second-order feedback path: when devices entered transient slowdowns (a GC cycle, an SLC eviction, a brief queue saturation), inferred costs rose, the rate limiter throttled tenants more aggressively, the offered load dropped, the device started looking healthier, and inferred costs fell again. The result was bursts of inverse correlation between offered load and observed limits. Static datasheet costs sidestep this entirely. The variance the static model misses is then absorbed by the two layers below it: the carry-forward credit pool (Section~\ref{sec:ratelimiter}) tolerates bounded under-accounting across quanta, and the distributed feedback controller (Section~\ref{sec:controller}) redistributes any structural under-utilization. The general principle is that in a layered scheduler, the lowest-level cost model should be \emph{stable} rather than \emph{accurate}; the layers above it can absorb modest inaccuracy but cannot absorb instability.

A second lesson concerns control-quantum granularity, where intuition turns out to be misleading. We initially assumed that the 200\,ms control quantum was a coarse upper bound and that finer-grained control (10 to 50\,ms) would be unambiguously better for tail latency. It is not. The per-quantum work---decoding tags, charging the cost-modeled token bucket, evaluating carry-forward state, and recomputing the slew-bounded extra-limit---is dominated by floating-point utilization math whose cost scales with the number of active consumer groups, which is in the thousands in our deployment. Halving the quantum doubles this overhead with no measurable improvement in SLO accuracy, because the rate-limiting decisions that materially affect tail latency happen \emph{within} the quantum at admission time, not at quantum boundaries. We settled on 200\,ms after observing that quanta as long as 500\,ms allowed visible latency excursions for interactive database workloads, while quanta below 100\,ms produced CPU stalls under heavy multi-tenant fan-in. The takeaway is that in a multi-tenant scheduler with thousands of consumer groups, the control-loop period is set by the cost of doing the per-tenant accounting rather than by the latency target itself; the latency target is held by admission-time pricing.

A third lesson is that carry-forward credit must be bounded, and that the bound is more than a controller-stability detail---it surfaces as an SLA artifact visible to tenants. Our earliest rate limiter used unbounded carry-forward. The pathology that motivated the one-second clamp was a recurring incident in which an idle database accumulated hours of credit overnight; when its overnight maintenance job started, it consumed the bucket all at once, monopolized the disk for several minutes, and triggered cascading p99 alarms on co-located tenants. The fix is structurally simple, but the operational implication is more subtle. A tenant's permitted burst is now an explicit, bounded SLA artifact (one second of 100\% capacity), not an emergent behavior of the credit pool. Communicating this bound clearly to application owners proved as important as enforcing it. The general lesson is that anti-windup in scheduler accounting is not just a controller-stability concern, as it usually is in classical control: it directly defines a tenant-visible burst SLA, and that SLA should be designed deliberately rather than inherited from an unbounded credit pool.

A fourth lesson, and the one that informs our most conservative design choices, is that distributed redistribution has to be slow. The temptation in a feedback controller is always to close error gaps quickly. In single-node control this instinct is correct; in cluster-wide control it is the most common cause of fabric-level instability. With higher gain or wider slew bounds, $N$ storage nodes that simultaneously observe ``my peers are idle, I can borrow their credits'' will all simultaneously raise their local ceilings and dump traffic onto the same RDMA fabric, producing a thundering herd that immediately re-collapses the steady state. The 5\% slew bound and 0.5 proportional gain we use are conservative for a reason: the controller has to behave as if it does not know what its peers are doing this tick, because in a strict shared-nothing cluster it does not. The general lesson is that a small-signal local controller, run independently on $N$ nodes, becomes a large-signal distributed controller whose bounds must be tuned for the worst-case correlated decision rather than for the typical local case.

A fifth and final lesson is one we did not initially appreciate: hardware transparency comes with a small ongoing engineering cost, and it is worth paying. Datasheet-derived costs require a per-generation lookup table that has to be updated whenever new hardware is introduced. This is the most frequently cited objection to the design from internal audiences. In practice, it is a once-per-generation pull request to a configuration file, set against a benefit that compounds across every tenant in every cluster: a tenant's provisioned limit means the same thing on day one and on hardware refresh day five years later. Operators and customers consistently rated this stability as worth the maintenance overhead. We think this generalizes to any system that abstracts physical capability: predictable performance semantics across hardware refreshes is itself a feature, and worth a small ongoing engineering cost.

Although we describe IORM in the context of a single platform, none of the three mechanisms are tied to that platform. The cost modeler relies only on per-generation datasheet numbers, which any disaggregated storage operator already tracks. The rate limiter is a standard quantum-based token bucket with anti-windup; the only thing unusual about it is that it is priced in normalized utilization rather than raw operations. The feedback controller is loosely coordinated: each node only needs an aggregate view of cluster-wide utilization for the tenants that have data on it, which is well within reach of any standard cloud-control plane. We expect the design to port to any disaggregated storage system that gives tenants a global IOPS budget, stripes data across heterogeneous storage nodes, and sees workloads with access skew. That covers most modern cloud block-storage services.

%===============================================================================
% SECTION 7 — RELATED WORK
%===============================================================================
\section{Related Work}\label{sec:related}

I/O scheduling and resource governance have been studied extensively across several system layers. We position IORM against four categories of prior work.

Kernel-level block schedulers such as CFQ~\cite{axboe2004linux}, BFQ~\cite{valente2012bfq}, and Kyber operate on a single device within a single host. They measure I/O cost via wall-clock busy time, which couples scheduling decisions to the physical characteristics of the underlying media. This makes them unsuitable for heterogeneous clusters where the same logical operation has different physical costs on different hardware generations. Furthermore, they have no mechanism for coordinating limits across multiple storage nodes.

Hypervisor and distributed schedulers such as mClock~\cite{gulati2010mclock} and PARDA~\cite{gulati2009parda} enforce proportional-share fairness across VMs or hypervisors. mClock provides per-device proportional-share scheduling with throughput and latency guarantees, but assumes a single shared device. PARDA distributes token-bucket enforcement across multiple hypervisors, but enforces static per-node limits and cannot dynamically redistribute unused entitlements when access patterns skew to a subset of nodes. Neither system addresses hardware-generation heterogeneity in I/O costing.

Network-layer governance systems such as IOFlow~\cite{ioflow2013} and Pisces~\cite{shires2016pisces} enforce policies along the storage I/O path at the network level. IOFlow provides a software-defined storage architecture with centralized policy enforcement, while Pisces enforces per-tenant bandwidth limits. However, both treat all storage bytes as equivalent, without modeling the physical cost differences between small random I/Os and large sequential I/Os, or between different media generations. This leads to inaccurate accounting for workloads that mix I/O sizes and types.

A separate strand of work addresses end-to-end SLO management for shared storage and co-located workloads. Cake~\cite{wang2012cake} schedules across multiple layers of the storage stack to enforce high-level latency SLOs in the presence of best-effort batch jobs; Heracles~\cite{lo2015heracles} co-locates latency-critical and best-effort workloads on the same hyperscale servers using coordinated control across cache, memory bandwidth, network, and CPU. These systems address the policy-enforcement problem at a higher layer than IORM and treat the underlying storage as homogeneous; they are complementary to a node-level governance mechanism that itself spans heterogeneous hardware.

Cloud-native block storage services such as AWS EBS~\cite{awsebsio} and OCI Block Volumes~\cite{ociblockperf} expose per-volume IOPS and throughput limits to tenants. While these services implement internal throttling, no published mechanism addresses cross-generation cost normalization or topology-aware credit redistribution within the storage cluster. Limits are enforced opaquely, and tenants have no visibility into whether their provisioned performance is affected by hardware heterogeneity or access skew.

The NVMe specification provides device-level isolation primitives such as NVM Sets and IO Determinism~\cite{nvmespec}, which allow firmware to partition physical resources within a single drive. These mechanisms provide per-device QoS but operate below the storage-server layer: they cannot enforce a cluster-wide IOPS budget spanning multiple drives and nodes, nor can they redistribute entitlements when access patterns shift across the cluster. IORM's scheduling operates at a higher level, treating device-level isolation as a complementary mechanism rather than a replacement.

Relative to this body of work, IORM is positioned by the combination of three capabilities that we are not aware of being present together in a single deployed system: datasheet-derived fixed-cost normalization that makes I/O accounting invariant across hardware generations, burst-governed quantum rate limiting that accommodates database micro-bursts without violating long-term SLOs, and a distributed adaptive feedback controller that dynamically redistributes unused credits across the cluster to resolve topological access skew. Each individual capability has analogs in prior work; what we report on is how they compose, and how that composition behaves in production---which is the contribution we believe is most useful to the cloud-engineering community.

%===============================================================================
% SECTION 8 — CONCLUSION
%===============================================================================
\section{Conclusion}

This paper presented IORM, a multi-stage distributed I/O scheduler for heterogeneous disaggregated storage. By decoupling logical accounting from physical latency via a hardware-aware cost modeler, IORM resolves the generational inconsistency inherent in mixed-hardware clusters. Its quantum-based rate limiter accommodates database micro-bursts while enforcing strict long-term limits, and its distributed feedback controller dynamically redistributes I/O credits to resolve topological access skew. Evaluation on an 8-node test cluster mirroring our production deployment demonstrates convergence to within 5\% of provisioned limits under extreme skew, scaling across 100 concurrent tenants without inter-tenant interference, and full throughput recovery within 15 seconds of a storage-node failure. Equally important, we have shared the operational lessons that shaped these mechanisms---why the cost model is deliberately static, why finer control quanta would not have helped, why anti-windup is an SLA-defining feature, and why a distributed controller must be slow. We hope these observations are useful to others building governance for the next generation of disaggregated cloud-storage systems.

%===============================================================================
% REFERENCES
%===============================================================================

\end{document}